# Poisson-type Multivariate Transfer Function Model Reveals Short-term Effects of Ambient Air Pollutants on Hospital Emergency room Visits for Cerebro-cardiovascular Diseases

**Abstract:** Laboratory experiments have shown that cardiovascular diseases are positively correlated to the concentration of ambient air pollutants, such as SO2, NO2, PM10, etc. It has also been repeatedly reported in many countries that increased concentration of ambient air pollutants leads to rise in hospital emergency room visitss for these diseases. These studies mainly adopt either regression analysis or preliminary models in time series analysis, while the multivariable transfer function model, a relatively newly developed model, has multiple advantages over the conventional linear regression on analyzing time series. This study attempts to quantify the association between concentrations ambient air pollutants and hospital emergency room visitss for cerebro-cardiovascular diseases in Beijing using a Poisson-type multivariate transfer function model. The results show that the RR values of SO2, NO2 and PM10 for a 50 g/m3 increase are 1.129, 1.092 and 1.069 respectively. The lags for the three pollutants are estimated to be 2 days, 1 day and 1 day respectively. Compared with the ambient pollutants, daily average temperature and relative humidity do not influence the daily count of hospital emergency room visits significantly.

## 1. Introduction

Hospital emergency room visits has been found to be associated with ambient air pollution in form of both gas and fine particulate matter (PM) in numerous studies [1-6], though the underlying pathological mechanism is not completely clear yet [7]. Most relevant studies focused on emergency room visitss due to respiratory disease, whereas the association between ambient air pollution and emergency room visitss for cerebrocardiovascular diseases does not seem to be sufficiently studied. Rudez [8] conducted laboratory experiments and verified that the incidence of cerebrocardiovascular diseases is positively correlated to the concentration of ambient air pollutants, such as $SO_2$, $NO_2$, $PM_{10}$ etc. It has also been reported that increased concentration of ambient air pollutants leads to rise in mortality rate or hospital emergency room visitss for these diseases [9-13].

Mathematical modeling has been widely used in epidemiology and environmental science [14-17]. Traditional regression models, e.g. linear regression and Poisson regression, are usually sophisticated enough to quantify the effects of independent variables on a dependent variable if great care is taken, but almost unconquerable difficulty would arise in cases where the respective delayed effects of pollutants must be considered [18]

Time series, i.e. a sequence of data points measured typically at successive times spaced at uniform time intervals, contains the information of time naturally. However, regression analysis, even properly applied, will inevitably omit the effect of the occurrence order of the events. In another word, high concentrations of pollutants a period of time ago may have a strong impact on hospital emergency room visits counts now, or even in the future. This time delay may be caused either by the reluctance of the patient to pay a visit to the hospital, or the prolonged course of the diseases dependent on the type of the patients' diseases and their susceptibility [19].

Another major problem that haunts these models is that when the data are relatively discrete i.e., the daily counts of hospital emergency room visitss are small numbers, canonical regression analysis does not work as well in practice, either because the goodness of fit is poor or because the residual error is so large that the obtained model is unlikely to be statistically significant. Thus a novel statistical model with both outstanding versatility and acceptable robustness must be utilized.

In this paper we introduce a Poisson-style multivariate transfer function model and apply it to evaluating the effects of pollutants on daily emergency room visitss for cerebrocardiovascular diseases. The transfer function model, which is also known as Auto-Regressive Integrated Moving Average with eXternal inputs (ARIMAX) was first developed by G. Box and G.M Jenkins in 1970. Ever since its emergence, the model has been widely used in fields of economics and engineering for its outstanding performance in forecasting and system controlling [20]. In 1990s, the model was introduced to the field of environmental medicine and public health to measure the effect of air pollutants on health [11]. Here we improved upon the classic ARIMAX model to develop a novel relative risk model to quantify the short-term effects of ambient pollutants with lag structures.

## 2. Materials and Methods

### 2.1. Data Description

Data on the day-to-day counts of hospital emergency room visits to Peking University 3rd hospital in urban Beijing for cerebro-cardiovascular diseases were obtained for the period January 1st 2004 to December 31st 2006 (1096 days in total). The cerebrocardiovascular diseases were divided into four categories which have been recorded separately: arrhythmia, heart failure, cerebrovascular diseases and others. Table1 shows average and standard deviation of daily counts categorized by different diseases. Close observation of the fluctuation of total daily counts (denoted as variable **Y** in the model) reveals that this time series is stochastic without any obvious increasing/decreasing trend or seasonal periodicity (Figure 1(F)).

Provided by Beijing Meteorological Bureau, concurrent data on ambient levels of sulfur dioxide (**$SO_2$**), nitrogen dioxide (**$NO_2$**) and **$PM_{10}$** (in milligram/m$^3$), average temperature (**avtemp**, in degree Celsius), and average relative humidity (**humidty**, in percent) on a daily basis are fed to the model as inputs series (Figure 1(A-E)).

**Table 1.** Average and Standard Deviation of Daily Hospital Emergency room visitss (Unit: person)

| diseases | mean | SD |
|---|---|---|
| Coronary Artery Disease | 1.5456 | 1.3992 |
| Arrhythmia | 1.2208 | 1.1549 |
| Heart Failure | 0.3568 | 0.6277 |
| Cerebrovascular Diseases | 3.3257 | 2.2126 |
| Others | 2.6688 | 2.0364 |
| Sum | 9.1177 | 3.8724 |

**Figure 2.** Profiles of air pollutants concentration (A, B and C), daily average temperature (D), daily relative humidity (E), and hospital emergency room visitss for cerebrocardiovascular diseases (F) during the study period.

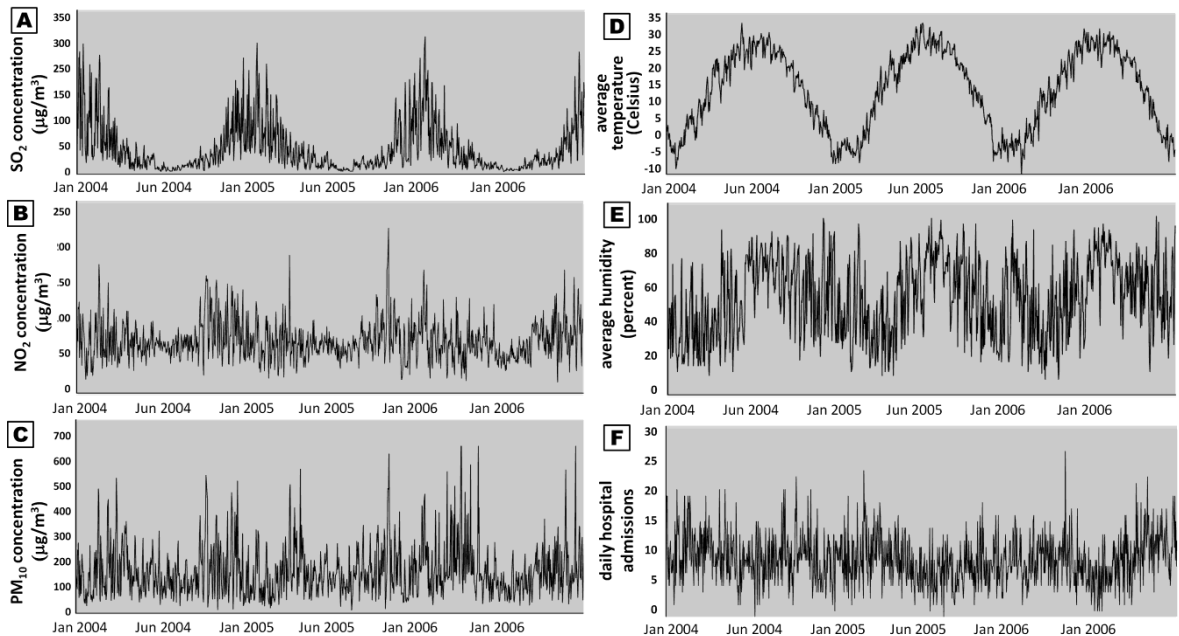

## 2.2. The canonical multivariate ARIMAX model

Suppose we have continuous real functions X(t) and Y(t) of time t, where X(t) is an independent variable and the following two assumptions are applied:

a. If X(t) remains constant for a enough long period of time, Y(t) is proportional to X(t), that is, Y(t)=gX(t), where the constant g is called steady-state gain that represents the impact on Y when X(t) are held constant over time;

b. If X(t) keeps fluctuating, at moment t, the changing rate of Y(t) is positively proportional to the difference of Y(t) and gX(t).

Based on the two reasonable assumptions, we develop the ordinary differential equation:

$$\frac{dY(t)}{dt} = \frac{1}{T}[gX(t) - Y(t)]$$

or

$$(1 + TD)Y(t) = gX(t) \qquad (1)$$

where T is a constant, the operator $D = d/dt$.

In case of time delay presence in such a system, the variation of X(t) cannot influence Y(t) immediately, but with a certain lag structure. Suppose $\tau$ denotes the duration of time delay, we thus change Equation (1) into:

$$(1+TD)Y(t) = gX(t-\tau) \qquad (2)$$

Similarly, for a second order system, where X(t) impacts upon $Y_2(t)$ by means of influencing $Y_1(t)$ first, the following two independent ODEs can be obtained:

$$(1+T_1 D)Y_1(t) = g_1 X(t-\tau) \qquad (3)$$

$$(1+T_2 D)Y_2(t) = g_2 Y_1(t-\tau) \qquad (4)$$

Combining (3) and (4) to eliminate $Y_1(t)$:

$$\left[1+(T_1+T_2)D+(T_1 T_2)D^2\right]Y_2(t) = g_1 g_2 X(t-\tau) \qquad (5)$$

In a more general form:

$$(1+\Xi_1 D+\Xi_2 D^2)Y(t) = gX(t-\tau) \qquad (6)$$

where $\Xi_1, \Xi_2$ are combinations of $T_1$ $T_2$, g is the combination of $g_1$ and $g_2$. For cases of even higher orders, e.g., order *R*, Equation (6) evolves into:

$$(1+\Xi_1 D+...+\Xi_R D^R)Y(t) = gX(t-\tau) \qquad (7)$$

We further consider the case where Y(t) is not only decided by X(t), but also by its up to *S*th order of derivatives:

$$(1+\Xi_1 D+...+\Xi_R D^R)Y(t) = g(1+H_1 D+...+H_s D^s)X(t-\tau) \qquad (8)$$

Until now, the model still only describes continuous functions X(t) and Y(t), and thus cannot be put into use in time series analysis wherein data are collected at certain moments with equal time intervals. Suppose the time interval is $\Delta t$. If the $\Delta t$ is small enough compared to the whole time scale:

$$DX(t) = \lim_{\Delta t \to 0} \frac{X(t)-X(t-\Delta t)}{\Delta t} \approx \frac{X_t - X_{t-1}}{t-(t-1)} = X_t - X_{t-1} = \nabla X \qquad (9)$$

Substitute Equation (9) into (8) to generate:

$$(1+\xi_1 \nabla+...+\xi_r \nabla^r)Y_t = g(1+\eta_1 \nabla+...+\eta_s \nabla^s)X_{t-b} \qquad (10)$$

where $\xi_i, \eta_i$ are constants and $b = \tau$.

If we further define the backward shifting operator:

$$B = 1-\nabla \qquad (11)$$

Equation (10) will be transformed into:

$$(1-\delta_1 B-...-\delta_r B^r)Y_t = (\omega_0 - \omega_1 B-...-\omega_s B^s)X_{t-b} \qquad (12)$$

where $\delta_i$ and $\omega_i$ are both parameters.

The simplified form of (12) is

$$Y_t = \delta^{-1}(B)\omega(B)X_{t-b} \qquad (13)$$

Adding the error term $N_t$ to Equation (13) would generate the final form:

$$Y_t = \delta^{-1}(B)\omega(B)X_{t-b} + N_t \qquad (14)$$

where $\delta^{-1}(B)\omega(B) = \dfrac{\omega_0 - \omega_1 B - ... - \omega_s B^s}{1 - \delta_1 B - ... - \delta_r B^r}$ and

$$N_t = \frac{\omega_0 - \omega_1 B - \ldots - \omega_{s_0} B^{s_0}}{1 - \delta_1 B - \ldots - \delta_{r_0} B^{r_0}} a_t \quad (a_t \text{ is white noise term}).$$

When the shifting operator $B$ is cancelled, the parameter

$$g = \frac{\omega_0 - \omega_1 - \ldots - \omega_s}{1 - \delta_1 - \ldots - \delta_r} \quad (15)$$

becomes the steady-state gain of $X_{t-b}$, which quantifies the contribution of $X_t$ to $Y_t$. If $n$ ($n>1$) time series can be used to predict the same time series $Y_t$ simultaneously, Equation (14) will be generalized into the so-called multivariate ARIMAX model:

$$Y_t = N_t + \delta_1^{-1}(B)\omega_1(B)X_{t-b_1} + \delta_2^{-1}(B)\omega_2(B)U_{t-b_2} + \delta_3^{-1}(B)\omega_3(B)V_{t-b_3} + \ldots$$

or more simply:

$$Y_t = N_t + \sum_{i=1}^{n} \delta_i^{-1}(B)\omega_i(B)X_{i,t-b_i} \quad (16)$$

*2.3. The Poisson-type multivariate transfer function model*

Since only a small portion of the population are admitted to hospital each day, the daily hospital emergency room visits is considered to be a random variable. It is also safe to assume that the hospital emergency room visitss on any two different days are independent from each other, which suggests that the underlying mechanism is Poisson process [18], and hence the probability of having Y hospital emergency room visitss on a certain day is expressed as

$$P(Y | \lambda) = \frac{e^{-\lambda}\lambda^Y}{Y!} \quad (17)$$

where $\lambda = E(Y)$ is the expectation of hospital emergency room visits on this day. For a non-homogeneous Poisson process, however, $\lambda$ evolves over time instead of remaining constant.

Poisson regression analysis is based on the assumption that the response variable *Y* has a Poisson distribution, and that the logarithm of its expected value, *log(E(Y))*, can be modeled by a linear combination of several predictor variables. Equation (15) was then revised as

$$\log(E(Y_t)) = N_t + \sum_{i=1}^{n} \delta_i^{-1}(B)\omega_i(B)X_{i,t-b_i} \quad (18)$$

*2.4. Parameter estimation*

The parameters *d, w* and *b* in Equation (17) were estimated using maximum likelihood method. Suppose we now have a data set of *m* input vectors $x_i \in \mathbb{R}^n, i = 1,2, \ldots m$ and a set of observed output values $y_i \in \mathbb{R}^1, i = 1,2, \ldots m$, according to the Poisson distribution's probability mass function given by Equation (16), the likelihood function in terms of the parameters *d, w* and *b* is

$$L(\delta, \omega, b | X, Y) =$$

$$\prod_{i=1}^{m} \frac{\exp[y_i(N_t + \sum_{i=1}^{n} \delta_i^{-1}(B)\omega_i(B)X_{i,t-b_i}) \times \exp(-N_t - \sum_{i=1}^{n} \delta_i^{-1}(B)\omega_i(B)X_{i,t-b_i})]}{y_i!} \quad (19)$$

Then parameter estimation could be accomplished by finding the set of parameters that allow the likelihood function to attain its maximum. Note that the classical least squares method can also be used for parameter estimation, and that there is no major difference between the two approaches except the heteroscedasticity introduced by the maximum likelihood method which is negligible within a narrow range of expected values [18].

*2.5. Modeling Procedures*

We now expect to apply this model to quantify the effects of the three pollutants: concentrations of $NO_2$, $SO_2$ and $PM_{10}$, as well as that of daily average temperature and relative humidity, on daily hospital emergency room visits with the available data. In order to efficiently establish the intricate statistical model, we made use of *SAS*® (Statistical Analysis System, SAS Institute Inc.), a versatile integrated system of software product. We also established a standardized model-building protocol (Figure (2)), for transfer function models with more than one input variable could be difficult to identify even for programmers with rich experience.

**Figure 2.** Flowchart of the model-building protocol

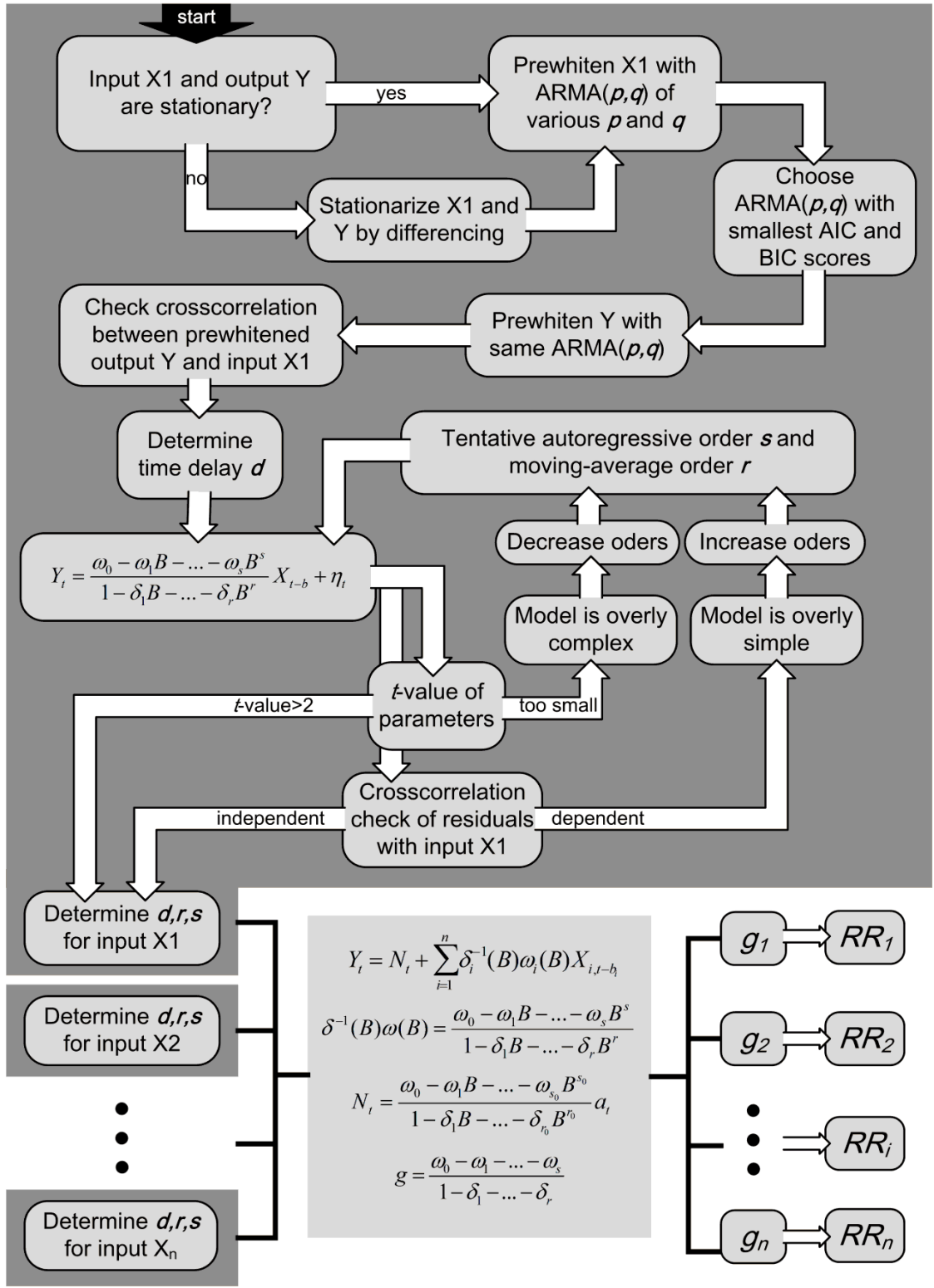

The first step is to identify an ARMA (autoregressive moving average) model describing the input time series and prewhiten the output series with the same model. The thumb's rule is that normally, ARMA(2,2) would be sophisticated enough to fit most time series [20]. In order to obtain the optimal model structure (sophisticated enough but not redundantly complicated), we always compare the AIC (Akaike's Information Criterion) and BIC (Bayesian Information Criterion) of ARMA(1,1), ARMA(1,2), ARMA(2,1), ARMA(2,2), ARMA(2,3), ARMA(3,2) and ARMA(3,3). The model with the lowest score of AIC and/or BIC is considered to be well-balanced between goodness of fit and model complexity [21].

Then the sample cross-correlation function between the prewhitened input and output time series should be calculated to determine the time delay between these two time series. The sample cross-correlation is a measure of the linear relationship between $Y_t$ and $X_{t-b}$ with varying lag-times $b$. If a spike in the cross-correlation at lag $\tau$ is detected, then we have $b=\tau$.

Next, the autoregressive order and moving-average order are also identified through trial and error to identify the structure transfer function, where two principles are applied: 1) if the estimated parameter $\omega$ or $\delta$ has an absolute $t$-value less than 2, it should be eliminated from the model to reduce model complexity; 2) if the resulted residuals are found to be correlated (dependent) to the input series, then either the autoregressive order or the moving-average order should be increased to enhance goodness of fit.

After the aforementioned three steps are applied to each of the input time series respectively, the overall structure of the multivariate ARIMAX model will be successfully generated and the steady-state gain of each input time series can then be obtained by parameter estimation described in Section 2.4.

### 2.6. Calculation of Relative Risk

Relative risk (RR), a ratio of the probability of the event occurring in the exposed group versus that in a non-exposed group [22], has been widely used as a measurement of the hazard of ambient pollutants. In order to calculate the RR value of each input variable, we approximate the input-output system with the steady-state gain $g_i$ by combining Equation (15) and (18):

$$E(Y_t) = \exp(N_t + \sum_{i=1}^{n} g_i X_{i,t-b_i}) \quad (20)$$

Thus for an increase of $\Delta x$ for the predictor variable $X_j, 1 \leq j \leq n$, the RR value was calculated as

$$RR = \frac{E(Y_t')}{population} \bigg/ \frac{E(Y_t)}{population} = \frac{E(Y_t')}{E(Y_t)}$$

$$= \frac{\exp(N_t + \sum_{i=1}^{n} g_i X_{i,t-b_i} + g_j \Delta x)}{\exp(N_t + \sum_{i=1}^{n} g_i X_{i,t-b_i})} = \exp(g_j \Delta x) \quad (21)$$

In order to compare the results with previous reports, all the RR values of ambient pollutant calculated in this study are for a 50 μg/m$^3$ increase in the pollutants, i.e. $\Delta x = 50$.

## 3. Results and Discussion

Seven ARMA models with different moving-average and autoregressive orders were built to prewhiten each input series, and their AIC and BIC values were calculated and compared to determine optimal orders (Table 2).

**Table 2.** ARMA models for the input time series

| model | SO$_2$ | | NO$_2$ | | PM$_{10}$ | | avtemp | | humidity | |
|---|---|---|---|---|---|---|---|---|---|---|
| | AIC | BIC | AIC | BIC | AIC | BIC | AIC | BIC | AIC | BIC |
| ARMA(1,1) | -4189 | -4174 | -5272 | -5257 | -2370 | -2355 | 9889 | 9904 | 9042 | 9057 |
| ARMA(1,2) | **-4348** | **-4328** | **-5295** | **-5275** | -2377 | -2357 | **9806** | **9826** | 9006 | 9026 |
| ARMA(2,1) | -4183 | -4163 | -5278 | -5257 | -2376 | -2356 | 9824 | 9844 | 9043 | 9063 |
| ARMA(2,2) | -4346 | -4321 | -5297 | -5272 | **-2383** | **-2358** | 9802 | 9827 | **8985** | **9010** |
| ARMA(2,3) | -4344 | -4314 | -5295 | -5265 | -2381 | -2351 | 9804 | 9834 | 8986 | 9016 |
| ARMA(3,2) | -4344 | -4314 | -5296 | -5266 | -2381 | -2351 | 9804 | 9834 | 8985 | 9015 |
| ARMA(3,3) | -4356 | -4321 | -5294 | -5259 | -2379 | -2344 | 9804 | 9839 | 8985 | 9020 |

*underlined figures are smallest AIC and/or BIC

Then the output series **Y** was also respectively prewhitened by the chosen models to estimate the time delay $d$ for each input series. Sequentially, the structural parameters, i.e., moving-average order $r$ and autoregressive order $s$, were determined through the trial and error procedure (Table 3). Using these model structures, the complete multivariate ARIMAX model was finally achieved:

$$\log(Y) = 1.733 + \frac{1 - 0.9541}{1 - 0.8732} \alpha_t + \frac{-0.3696 + 0.4723B}{1 - 0.0772B - 0.8806B^2} \times 10^{-3} \cdot (SO_2)_{t-2}$$
$$+ \frac{3.011}{1 + 0.7138B} \times 10^{-3} \cdot (NO_2)_{t-1} + \frac{0.2257 - 0.4513B + 0.3605B^2}{1 - 0.8992B} \times 10^{-3} \cdot (PM10)_{t-1}$$
$$- \frac{0.0006}{1 - 0.1398B - 0.2337B^2} \cdot (avtemp)_t + \frac{-0.003 - 0.00154B + 0.00187B^2}{1 + 0.8444B} \cdot (humidity)_{t-1} \quad (22)$$

The RR values of each pollutant as well as the daily average temperature and relative humidity were calculated and listed in Table 3.

**Table 3.** Structural parameters and estimated RR value of input series

| parameters | $SO_2$ | $NO_2$ | $PM_{10}$ | avtemp | humidity |
|---|---|---|---|---|---|
| time delay: $d$ (days) | 2 | 1 | 1 | 0 | 1 |
| moving average order: $r$ | 2 | 1 | 1 | 2 | 1 |
| autoregressive order: $s$ | 1 | 0 | 2 | 0 | 2 |
| $g\Delta x$ | 0.1217 | 0.0878 | 0.0669 | -0.0096[a] | -0.0145[b] |
| RR | 1.129 | 1.092 | 1.069 | 0.9905[a] | 0.9856[b] |

[a] for an increase of 10 degrees Celsius, i.e. $\Delta x = 10$.
[b] for an increase of 10 per cents, i.e. $\Delta x = 10$.

The results show that the RR values of $SO_2$, $NO_2$ and $PM_{10}$ are 1.129, 1.092, and 1.069 respectively, which suggests that the levels of all the three ambient pollutants are hazardous factors for cerebrocardiovascular diseases, as all their RR values are bigger than 1. Sunyer et al [23] reported that adults asthma emergency room visitss in four cities in Europe were positively associated with $SO_2$, $NO_2$ and $PM_{10}$ (black smoke), with RR=1.012, 1.078 and 1.032 for a 50 μg/m$^3$ increase respectively. Spix et al [10] conducted a massive meta analysis in 5 cities in Europe and concluded that the RR values of these three pollutants on daily respiratory emergency room visits were 1.009, 1.010 and 1.010 respectively. In comparison with these results, the RR values of the three pollutants obtained in this study are significantly larger, suggesting that cerebrocardiovascular emergency room visitss may be more significantly positively associated with these pollutants than respiratory emergency room visitss.

Previous studies also estimated the time delay of the short-term effect of the pollutants. Bates and Sizto [2] reported a 2-day, 2-day and 1-day lag for $SO_2$, $NO_2$ and $PM_{10}$ respectively, whereas Sunyer et al's [23] results are 2-day, 1-day and 0-day respectively. The lags generated by the multivariate transfer function model are similar to those studies, showing the favorable reliability of the model in modeling structure identification and parameter estimation. It can also imply that the ambient pollutants might give rise to cerebrocardiovascular and respiratory diseases through similar or common physiological mechanisms.

Table 3 also shows that daily average temperature and relative humidity do not influence the daily count of hospital emergency room visits significantly, as the RR values of both of them are fairly close to 1. One study of Spix et al [10] purposely estimated the hazard of ambient pollutants in cold and warm seasons respectively and detected no significant difference between their RR values. Both their study and ours suggest that the short-term effect of temperature is negligible compared with ambient pollutants.

Though both multivariate transfer function model and canonical linear regression model can be used to explain or predict one variable with several independent variables, the former works in a more generalized and sophisticated way. If we eliminate the chronological information contained in time series, i.e, the data were collected without sequential order, the transfer function model will degenerate into ordinary regression analysis. In this sense, the transfer function model has taken a great leap forward by expanding linear regression along the time axis. Therefore, the meaningful extra time dimension would have been lost if the data are processed with linear regression instead. This relatively newly developed model is not only sophisticated enough to measure the RR value of each pollutant, but also able to analyze and quantify the delay effect of them [20]. An additional merit of this model is that it is considerably tolerant to the discreteness of available data, which fails logistic regression and path analysis.

## 4. Conclusions

This study attempts to quantify the association between concentrations of the ambient air pollutants and hospital emergency room visitss for cerebrocardiovascular diseases in Beijing using a Poisson-type multivariate transfer function model. The results show that the RR values of $SO_2$, $NO_2$ and $PM_{10}$ for a 50 μg/m$^3$ increase are 1.129, 1.092 and 1.069 respectively, significantly higher than the RR values of these pollutants on hospital emergency room visits for respiratory diseases. The lags for the three pollutants are estimated to be 2 days, 1 day and 1 day respectively. Compared with the ambient pollutants, daily average temperature and relative humidity do not influence the daily count of hospital emergency room visits significantly. This study demonstrates that the Poisson-type multivariate transfer function model can be successfully applied in modeling intricate associations between the output time series and multiple input time series.

**Conflict of Interest**

The authors declare no conflict of interest.

**References**

1   Bates D.; Baker-anderson M.; Sizto R. Asthma attack periodicity: a study of hospital emergency visits in Vancouver. *Environ Res* **1990,** 451, 51-70.
2   Bates D.; Sizto R. Air pollution and hospital emergency room visitss in southern Ontario: the acid summer haze effect. *Environ Res* **1987,** 43, 317-31.